\documentstyle[preprint,eqsecnum,aps]{revtex}

\begin{document}

\title{A lattice fermion without doubling}
\draft

\author{A. Hayashi, T. Hashimoto, 
    M. Horibe${}^{*}$ and H. Yamamoto}

\address{
    Department of Applied Physics, 
    Fukui University, Fukui 910\\
    ${}^*$Department of Physics, Faculty of Education, Fukui University\\
    Fukui 910
}

\date{\today}
\maketitle

\begin{abstract}

A free fermion without doubler is formulated on 1+D dimensional
discrete Minkowski space-time.  The action is not hermitian but
causes no harm.  In 1+3 dimensional massless case
the equation describes a single species of Dirac particle 
in the continuous space-time limit.  In 1+1 dimensional 
massless case the 
equation is the same as the automaton equation by 't Hooft 
and describes a chiral fermion.
The time evolution operator is unitary and the norm is conserved.
For interacting fermions with gauge fields the evolution operator
is not unitary. If it is considered as an approximation for the
theory on continuous space-time, the path integral formalism can 
be applied, where the fermion without doubling is used.
Consequences of loosening the unitarity condition on the time 
evolution operator is discussed.
\end{abstract}

\section{Introduction}

As is well known, we necessarily meet the difficulty of so-called fermion 
doubling when we formulate a fermion on lattice space-time. 
Nielsen-Ninomiya proved under certain assumptions that a chiral 
fermion cannot be formulated on Kogut-Susskind lattice 
(=continuous time and discrete space) \cite{nielsen}.
At present it is 
widely believed that fermion doubling is inevitable on discrete 
space-time in general.   In order to get rid of this unwanted 
partner of the fermion Wilson introduced an additional term in 
Lagrangian to make the mass of the partner very large \cite{wilson}.
Hence the partner has 
no effect at low energy.  Another attempt is one by 
Susskind \cite{susskind}.  
He showed these two fermions could be interpreted as isodoublets
(=fermions of two different flavours).
  
In the present paper we want to show that a fermion can be 
formulated without doubling on Minkowski lattice space-time (Sec.2). 
In order to avoid the no-go theorem we do not assume the hermiticity 
of action which was one of the assumptions of Nielsen-Ninomiya. 
The non-hermitian action may lead to some problems, but in our 
case it seems not so serious.  For example, the action and the
hermitian conjugate action yield two consistent field equations. 
This means both the real and imaginary parts of the action take 
stationary values at the same time, if the equation is satisfied.
However, it is not yet clear whether the non-hermiticity of
the action produces further difficulties.  

    The equation we obtained on 1+D dimensional discrete Minkowski
space-time has no fermion doubling solution. 
In 1+3 dimensions it tends to Dirac equation in the continuous limit.
The time evolution operator is unitary and therefore, the norm 
is conserved.  In 1+1 dimensions a chiral fermion can be described
in massless case which is the same as the automaton model 
of 't Hooft \cite{thooft}.  However, in 1+3 dimensions a 
completely chiral
fermion cannot be described as in the case of Susskind.
This situation is not improved even if the unitarity of 
time evolution operator is not assumed (Sec.4). 

    In the above mentioned we treated only a free fermion field. 
In the case of interacting fields, however, the situation is
completely different.  It is difficult to assume the unitarity 
of the time evolution operator. If we stand on the viewpoint that 
the discrete space-time is an approximation of continuous case,
we may disregard this point, because the unitarity of
time evolution operator is recovered in the continuous 
space-time limit.  In the lattice gauge theory which is 
formulated on Euclidean space-time, the unitarity of time 
evolution operator is not considered seriously.  
Using the method of path integral formalism we can calculate
any quantities in principle without the effect of fermion 
doubling, though the non-hermiticity might bring some 
practical difficulties (Sec.3).  


\def\bx{{\bf x}}

\section{Non-interacting fermions}

    In this section we develop a lattice theory of non-interacting
fermions without doublers in $1+D$ dimensional discrete space-time.
We take a hyper-rectangular lattice with lattice constants $\tau$ 
and 
$\sigma$ for the time and space direction, respectively. 
Lattice points are represented by 
a set of $1+D$ integers 
$(t,x^1,\ldots,x^D)\equiv(t,\bx)\equiv x$.
At each site is attached a spinor variable $\Psi(x)$, whose 
number of components is not specified for the moment. 

\subsection{Equations of motion}

    We assume that equations of motion for $\Psi$ take the form:
\begin{equation}
  \Psi(t+1,\bx)=U\Psi(t,\bx),
 \label{eq}
\end{equation}
where $U$ is the time evolution operator acting on the $\bx$ and 
spinor space. Postulating some properties on $U$,
we will determine $U$ and the minimal dimension of the spinor 
space for $\Psi$. 

    The time evolution operator $U$ is assumed to be linear in 
the unit shift operators in the $\bx$-space so that only the nearest
 neighbor sites are 
coupled in the action, which will be given in the next subsection.
Thus we have 
\begin{equation}
   U=\sum^D_{i=1} \left( A_i S_i + B_i S_i^\dagger \right) + C,
   \label{U}
\end{equation}
where the unit shift operators $S_i$'s are defined by
\begin{equation}
     S_i \Psi(t,x^1,\ldots,x^i,\ldots,x^D)
   =     \Psi(t,x^1,\ldots,x^i+1,\ldots,x^D),
\end{equation}
and $A_i$, $B_i$ and $C$ are matrices with respect to spinor
indices.

    We furthermore assume that the time evolution operator 
is unitary:
\begin{equation}
  UU^\dagger=1.
  \label{unitary}
\end{equation}
 This condition is necessary for 
the theory to be well-defined quantum theory on discrete time. 
However, if the theory is considered to be an approximation to 
the continuum theory,
we could loosen the condition in such a way that the 
unitarity should be  recovered only in the continuum limit. 
In this section we assume the unitarity, and 
consequences of 
loosening the condition will be discussed in Sec. 4.

    We require that each component of $\Psi$ satisfies the 
discrete version of the Klein-Gordon equation:
\begin{equation}
 {U-2+U^{-1} \over \tau^2}
- \sum^D_{i=1}\, { S_i -2 + S_i^\dagger \over \sigma^2} 
   + M^2 = 0,
 \label{Klein}
\end{equation}
where M is the hermitian mass matrix.
The dispersion relation implied by this 
equation is given by  
\begin{equation}
   {4 \over \tau^2}\sin^2 {k^0 \over 2}
 - {4 \over \sigma^2}\sum_{i=1}^D \,\sin^2{k^i \over 2}
 - M^2  = 0,
 \label{disp}
\end{equation}
where $k^0$ and ${\bf k}$ are introduced through the Fourier
transform:
\begin{equation}
  \Psi(t,\bx) = \int_{-\pi}^\pi dk^0d{\bf k}
      \, e^{-i(k^0t-{\bf k}\bx)}\,\Psi_{k^0\,{\bf k}}.
\end{equation}
Evidently there is no doubling problem with
Eq. (\ref{disp}), which has also been pointed 
out by Yamamoto\cite{yamamoto}.
Therefore, it is this equation (\ref{Klein}) which assures
 that no doublers
appear in our formalism.
In the following we assume $M=0$ and at the end of this subsection
we will discuss the case of $M\ne0$.

    Substituting $U$ given by Eq. (\ref{U}) in the Klein-Gordon 
equation (\ref{Klein}), where $U^{-1}$ is replaced by
$U^\dagger$, we find the relations:
\begin{eqnarray}
   & & A_i+B_i^\dagger=r,  \nonumber \\
   & & A_i^\dagger+B_i=r,  \nonumber \\ 
   & & C+C^\dagger-2=-2D,
 \label{kg-condition}
\end{eqnarray}
where $r$ is defined as $(\tau /  \sigma)^2$. 

    On the other hand the unitarity condition (\ref{unitary}) 
requires that
\begin{eqnarray}
    & & A_iB_j^\dagger+A_jB_i^\dagger=0,  \nonumber \\
    & & A_iA_j^\dagger+B_jB_i^\dagger=0,\ \ \ \ \ \ \ \ \ \ \
                             ({\rm for}\ i\neq j)\nonumber \\
    & & A_iC^\dagger+CB_i^\dagger=0,      \nonumber \\
    & & \sum^D_{i=1}\left( A_iA_i^\dagger+B_iB_i^\dagger \right)
                 +CC^\dagger = 1.
  \label{u-condition}
\end{eqnarray} 

    Combining those relations (\ref{kg-condition}) and 
(\ref{u-condition}), we find that $U$ takes the following form:
\begin{equation}
   U = \sum^D_{i=1} 
  \left\{ 
      {r \over 2}(1+a_i) S_i + {r \over 2}(1-a_i^\dagger) S_i^\dagger 
  \right\}
        + (1-Dr) + iC_I,
\end{equation}
where 
the operators in the spinor space 
$a_i$'s and $C_I$ (the imaginary part of $C$) are yet to be 
determined by the following algebra:
\begin{eqnarray}
  & & \left\{ a_i , a_j \right\} = 2, \nonumber \\
  & & \left\{ a_i , C_I \right\} = -2i\,(1-Dr), \nonumber \\
  & & \left\{ a_i , a_j^\dagger \right\} = -2, \ \ \ \ \ \ \ \ \ \ \ \ \ 
              ({\rm for}\ i\neq j) \nonumber \\
  & & \sum^D_{i=1} \left\{ a_i , a_i^\dagger \right\} 
                           = D \left( {8 \over r} -4D -2 \right)
                               -{4 \over r^2}C_I^2 .
 \label{algebra}
\end{eqnarray}
Here a set of curly brackets $\{\ ,\ \}$ is an anticommutator.

    Using this algebra, we find that $C_I$ is not linearly independent 
of $a_i$ but expressed as
\begin{equation}
  C_I = i {r \over 2} \,\sum^D_{i=1} 
           \left( a_i - a_i^\dagger \right),
  \label{ci}
\end{equation}
which can be readily verified by showing the square of the
difference between both sides of the equation vanishes.
The relation (\ref{ci}) has an interesting consequence
on  the lattice constants $\tau$ and $\sigma$.
Calculating $C_I^2$ by Eq. (\ref{ci}), one finds
\begin{equation}
  C_I^2 = Dr\,(1-Dr),
  \label{ci2}
\end{equation}
which should be positive definite since $C_I$ is hermitian. 
Thus we have
\begin{equation}
  r\equiv \left({\tau \over \sigma}\right)^2 \leq {1 \over D}.
  \label{rD}
\end{equation}
Yamamoto has obtained this inequality as well as the more general
one mentioned below involving the finite mass\cite{yamamoto}.
The argument is that the $k^0$ should be real in the dispersion 
relation (\ref{disp}) 
, which is consistent with our requirement of 
the unitarity of $U$.

    It is convenient to decompose $a_i$ into the real and imaginary
part as $a_i=X_i+iY_i$, where $X_i$ and $Y_i$ are hermitian.
By Eqs. (\ref{ci}) and (\ref{ci2}) the algebra for 
$X_i$ and $Y_i$ is reduced to be
\begin{eqnarray}
  & & \left\{ X_i , X_j \right\} =2{1 \over r}\,\delta_{ij}, 
                                  \nonumber \\
  & & \left\{ X_i , Y_j \right\} = 0,          \nonumber    \\
  & & \left\{ Y_i , Y_j \right\} =
         2\left({1 \over r}\,\delta_{ij}-1\right).
\end{eqnarray}
Although the anticommutation relation between $X_i$'s is diagonal
with respect to the indices $i$ and $j$, the one between $Y_i$'s
is not.  The matrix $2(\delta_{ij}/r-1)$, however, can be easily
diagonalized. The eigenvalues are $2(1/r-D)$ for the 
eigenvector $(1,1,\ldots,1)$ and $2/r$ for any
vectors orthogonal to $(1,1,\ldots,1)$. Thus the
latter eigenvalue is $(D-1)$-fold degenerate.

    Therefore, when $r<1/D$, we have $2D$ non-zero hermitian 
operators
whose anticommutation relation is diagonal. Namely $X_i$ and 
$Y_i$, after the diagonalization, form the Clifford algebra 
up to trivial normalization factors:
\begin{eqnarray}
 & & \Gamma_i^\dagger = \Gamma_i,\ \ \ \ \ \ \ \ \ \ \ i=1,\ldots,N \nonumber \\
 & & \left\{ \Gamma_i, \Gamma_j \right\} = 2 \delta_{ij}
\end{eqnarray}
with $N$ being $2D$ in this case. So the dimension of the 
irreducible representation
for $X$ and $Y$ is $2^D$ and $\Psi(x)$ should have $2^D$
components at least.
  
    In the case of $r=1/D$, where $r$ takes the maximal value 
allowed by Eq. (\ref{rD}), one of the eigenvalues $2(1/r-D)$ 
vanishes and the linear relation $\sum Y_i = 0$ holds.
In this case we have the Clifford
algebra with $N=2D-1$, whose irreducible representation has the 
dimension of $2^{D-1}$. Consequently $\Psi(x)$ has 
$2^{D-1}$ components. 

    Assuming $r=1/D$, so that $\Psi(x)$ has 
the smaller number of components $2^{D-1}$, let us
investigate the explicit form of the time 
evolution operator $U$ for the cases $D=1$ and $D=3$.

\bigskip
\noindent
{\bf 1+1 dimensional case}
\bigskip

    $\Psi(t,x^1)$ is one-component field and $U$ is given by 
\begin{equation}
 U = {1 \over 2}(1+X_1)S_1 + {1 \over 2}(1-X_1)S_1^\dagger,
\end{equation}
where $X_1^2=1$.
When $X_1=1$, we have $U=S_1$ and the equation of motion is simply  
$\Psi(t+1,x^1)=\Psi(t,x^1+1)$, which clearly simulates without 
doublers the chiral left moving fermion 
$(\partial_0-\partial_1)\Psi=0$ in the $1+1$ continuum theory. 
It is interesting that the correspondence is established by 
replacing the time derivative $\partial_0\Psi$ by the forward
difference quotient but the spatial derivative $\partial_1\Psi$
by the backward one. 
This equation of motion has also been given by 't~Hooft et al. 
in the cellular automaton approach\cite{thooft}. 
In the same way, the case of $X_1=-1$ describes the right moving
chiral fermion without doublers. 

\bigskip
\noindent
{\bf 1+3 dimensional case}
\bigskip

    The ratio of lattice constants $\tau/\sigma= \sqrt{r}$
is given by $1/\sqrt{3}$ and  $\Psi(t,\bx)$ is 4-component field. 
The time evolution operator $U$ then takes the form:     
\begin{equation}
U  =  1+\sum_{i=1}^D\tau \Gamma_i{S_i-S_i^\dagger \over 2\sigma}
    + \sum_{i=1}^D{1 \over 6}(1+iY_i)(S_i-2+S_i^\dagger),  
\label{U4}
\end{equation}
where $Y_i$'s are given by
\begin{eqnarray}
& & Y_1 = \sqrt{2}\Gamma_4,   \nonumber \\
& & Y_2 = -\sqrt{{1 \over 2}}\Gamma_4 + \sqrt{{3 \over 2}}\Gamma_5,
                                   \nonumber  \\
& & Y_3 = -Y_1-Y_2.
\end{eqnarray}
More explicitly those $\Gamma_i$'s $(i=1,\ldots,5)$ are expressed by the
Dirac matrices:
\begin{equation}
  \Gamma_i = \gamma^0\gamma^i\  (i=1,2,3),\ \ \ \ \   
  \Gamma_4 = \gamma^0, \ \ \ \ \ \Gamma_5 = \gamma^5.
\end{equation}

    The second term in the time evolution operator (\ref{U4})
is reduced to be $\tau \gamma^0\gamma^i\partial_i$ in the 
continuum limit. The last term in $U$ goes to zero 
as $\tau^2$  in the 
na\"{\i}ve continuum limit, and it reminds us of the Wilson term except 
for the spinor operators $Y_i$ in front.  
It is this spinor dependence 
that eliminates doublers completely, whereas the Wilson term 
only gives doublers a large mass.  

   Now we briefly mention only the results for the case 
of $M \ne 0$, which is 
more involved. Instead of the algebra (\ref{algebra}), 
we have
\begin{eqnarray}
  & & \left\{ a_i , a_j \right\} = 2, \nonumber \\
  & & \left\{ a_i , C_I \right\} = -2i\,C_R, \nonumber \\
  & & \left\{ a_i , a_j^\dagger \right\} = -2, \ \ \ \ \ \ \ \ \ \ 
              ({\rm for}\ i\neq j) \nonumber \\
  & & \sum^D_{i=1} \left\{ a_i , a_i^\dagger \right\} 
                           = (1 - C^\dagger C){4 \over r^2} 
                               -2D ,
 \label{algebraM}
\end{eqnarray}
and the real part of $C$ is given by $C_R=(1-Dr)-\tau^2M^2/2$.
An interesting consequence is that $M^2$ commutes with $a_i$ and 
$C_I$, and therefore, $M^2$ is just a number in the irreducible 
representation. We can furthermore show that $r \leq 
(1-\tau^2 M^2/4)/D$ and even if the equality holds, $\Psi(x)$
has $2^D$ components at least.

\subsection{Action}

    We have shown that one can construct equations of motion
for $\Psi(x)$ without doublers by the time evolution 
operator $U$ of the form (\ref{U}), which is local and unitary. Now we construct the 
action on the lattice which leads to the equations of motion. 

    We immediately see that 
the desired equations of motion derive from 
the variation with respect to $\Psi^\dagger$ of the action:  
\begin{equation}
   S = i\sum_{t,\bx}\, \Psi^\dagger(t+k,\bx)
       \left( \Psi(t+1,\bx) - U\Psi(t,\bx) \right),
   \label{action}
\end{equation}
where $k$ is an arbitrary integer. The above action is evidently
local but not hermitian. It should be noted, however, that
the variation of the action with respect to $\Psi$ also gives 
the equivalent equations of motion, which is a consequence of 
the unitarity of $U$.

    As we have seen in the last subsection, the single chiral 
fermion can safely reside in 1+1 dimensional discrete 
space-time. 
This does not contradict the Nielsen-Ninomiya theorem, since
our action is local but not hermitian. 

    We will now show that the quantization through the path
integral with the action (\ref{action}) is equivalent
to the equations of motion (\ref{eq}) with the canonical
equal-time commutation relations 
$\{\hat\Psi(t,\bx),\hat\Psi^\dagger(t,\bx^\prime)\}
=\delta_{\bx\bx^\prime}$. Here and in what follows, $\hat\Psi$ 
and $\hat\Psi^\dagger$ should be understood to be canonically
quantized fields whereas $\Psi$ and $\Psi^\dagger$ are
Grassmannian variables. 

    First we introduce the 
time evolution operator $\hat U$ in the fermion Fock space,
which should not be confused with $U$, as follows:
\begin{equation}
  \hat U = \exp\left( -i\sum_\bx\hat\Psi^\dagger(t,\bx)
                      H\hat\Psi(t,\bx)
                \right), 
\end{equation}
where the hermitian operator $H$ in the $\bx$ and spinor 
space is defined through $U = \exp(-iH)$. The $\hat U$ is 
evidently unitary and reproduces the equations of motion:
\begin{equation}
   \hat\Psi(t+1,\bx)=\hat U^{-1} \hat\Psi(t,\bx) \hat U
                    =U \hat\Psi(t,\bx).
\end{equation}
We also find that $\hat U$ is independent of $t$.

    Now we calculate the matrix element of $\hat U$ between 
fermion coherent states $\mid \Psi(\cdot)>$ and 
$\mid \Psi^\prime(\cdot)>$. Here the coherent states are defined
as
\begin{equation}
    \mid \Psi(\cdot)> \equiv \exp
       \left( -\sum_\bx\Psi(\bx)\hat\Psi^\dagger(0,\bx) \right)
 \mid 0>, 
    \ \ \hat\Psi(0,\bx)\mid 0> = 0.
\end{equation}
The matrix element of $\hat U$ can be readily calculated as
\begin{equation}
      <\Psi^\prime(\cdot)\mid \hat U \mid \Psi(\cdot)> \\
  = \exp\left( \sum_\bx\Psi^{\prime\dagger}(\bx)U\Psi(\bx) \right).
\end{equation}

    Noting that the completeness relation of the coherent 
states is given by
\begin{equation}
 1  =  \int\prod_\bx 
              \left[d\Psi^\dagger(\bx) d\Psi(\bx)\right] 
     \exp\left( -\sum_\bx\Psi^\dagger(\bx)\Psi(\bx)\right)
     \mid \Psi(\cdot)><\Psi(\cdot) \mid,
\end{equation}
for the transition amplitude from the initial state 
$\mid \Psi(t_0,\cdot)>$ 
to the final state $ \mid \Psi(t_n,\cdot)>$ we have 
\begin{eqnarray}
 & & <\Psi(t_n,\cdot)\mid \hat U^n \mid \Psi(t_0,\cdot)> 
                                                   \nonumber  \\
 &=&\exp\left( \sum_\bx \Psi^\dagger(t_n,\bx)\Psi(t_n,\bx) \right) 
     \int{\cal D}\Psi^\dagger{\cal D}\Psi  
                                                   \nonumber  \\ 
 & &    \exp\left(-
          \sum_{t=t_0,\bx}^{t_{n-1}}
             \Psi^\dagger(t+1,\bx)
             \left(\Psi(t+1,\bx)-U\Psi(t,\bx)\right)
             \right),
\end{eqnarray}
which establishes the equivalence of 
the two ways of quantization after redefining $\Psi^\dagger(t+k,\bx)$ as 
$\Psi^\dagger(t+1,\bx)$ in 
the action (\ref{action}). 

    Finally we will give the explicit form of the propagator,
which is the inverse of the kernel in the action
(\ref{action}).
Defining the unit shift operator in the time direction as 
$S_0\Psi(t,\bx)=\Psi(t+1,\bx)$, we have
\begin{eqnarray}
  <T\Psi(x)\Psi^\dagger(x^\prime)>
 & =& \left( {1 \over 1-S_0^\dagger U} \right)_{x,x^\prime} 
                          \nonumber \\
 & =& \left(   
     { S_0 - U^\dagger  \over
       (S_0-2+S_0^\dagger)-(U-2+U^\dagger) }
    \right)_{x,x^\prime}.
\end{eqnarray} 
Evidently the propagator has no extra poles,
since in the momentum space
its denominator 
takes the form of the left hand side of the dispersion relation 
(\ref{disp}), due to the Klein-Gordon equation (\ref{Klein}). 


\section{Interaction with gauge fields}

We have seen in the previous section that the lattice fermion without
doubling can be formulated for the free case in 1+D dimensions.
In this section we consider the interaction of the fermion with
gauge fields.

The interaction is introduced by replacing the shift operators by covariant ones:

\begin{equation}
S_{0} \rightarrow S_{0}(x) \equiv V(x,x+\hat{0}) S_{0} ,
\end{equation}

\begin{equation}
S_{i} \rightarrow S_{i}(x) \equiv V(x,x+\hat{i}) S_{i} ,
\end{equation}

\noindent
where $\hat{0}$ and $\hat{i}$ are unit vectors along the time and
$i$-th space direction, respectively, and
$V(x,y)$ is a link variable connecting sites $x$ and $y$.
After these replacements the action is written as

\begin{equation}
S = i \sum_{x} \Psi^{\dagger}(x)
       \{ 1 - S_{0}^{\dagger}(x)U(x) \} \Psi(x) + S_{\rm gauge} ,
\end{equation}

\noindent
where \( U(x) \) is
defined by replacing $S_{i}$ with $S_{i}(x)$ in Eq. (\ref{U}) and
\( S_{\rm gauge} \) is the action for gauge fields.
It is easily seen that the action is invariant under the lattice
gauge transformation with a gauge group element \( g(x): \)

\begin{equation}
\Psi(x) \rightarrow g(x) \Psi(x) ,
\end{equation}

\begin{equation}
V(x,y) \rightarrow g(x) V(x,y) g(y)^{-1} .
\end{equation}

\noindent
It is also checked that the new action has the correct continuum
limit.

Using this new action and path integral formalism, we can
calculate the vacuum expectation
value of an arbitrary operator \( O \) by the formula:

\begin{equation}
<O> = \int {\cal D } \Psi^\dagger {\cal D } \Psi {\cal D } V
         \makebox[2mm]{} O \makebox[1mm]{} \exp ( iS )
         \makebox[2mm]{} / \int {\cal D } \Psi^\dagger {\cal D } \Psi {\cal D } V
         \exp ( iS ) .
\end{equation}

In the usual case the gauge interaction is introduced in an analogous way
using the symmetric discretization of time.
But the unitarity of the time evolution of fermion fields is not clear
for the symmetric discretization, since the field at time $ t+1 $ depends
on fields at $ t $ and $ t-1 $.
Usually this is not considered seriously because the unitarity
is recovered in the continuum limit.
However, in our case we have the explicit time evolution operator
for a finite lattice spacing,
and therefore, we can go a little further in the discussion of unitarity.

We assume the Klein-Gordon condition and the unitarity condition also in this
case.
The Klein-Gordon condition gives the same relations Eq. (\ref{kg-condition})
for \( A \) , \( B \) and \( C \)
as in the free case.
From the unitarity requirement we have the following relations :

\begin{eqnarray}
A_{i}B_{j}^{\dagger}V(x,x+\hat{i})V(x+\hat{i},x+\hat{i}+\hat{j})  \makebox[3cm]{} \nonumber \\
+ A_{j}B_{i}^{\dagger}V(x,x+\hat{j})V(x+\hat{j},x+\hat{i}+\hat{j}) = 0 ,
\end{eqnarray}

\begin{eqnarray}
\hspace*{10mm}
A_{i}A_{j}^{\dagger}V(x,x+\hat{i})V(x+\hat{i},x+\hat{i}-\hat{j}) \makebox[40mm]{} \nonumber \\
+ B_{j}B_{i}^{\dagger}V(x,x-\hat{j})V(x-\hat{j},x+\hat{i}-\hat{j}) = 0 , \makebox[5mm]{} ( i \ne j 
)
\end{eqnarray}

\begin{equation}
A_{i}C^{\dagger} + CB_{i}^{\dagger} = 0 ,
\end{equation}

\begin{equation}
\sum_{i=1}^{D} ( A_{i}A_{i}^{\dagger} + B_{i}B_{i}^{\dagger} )
      + CC^{\dagger} = 1 ,
\end{equation}
which correspond to Eq. (\ref{u-condition}) in the free case.

The last two relations are the same as before, but the first two contain link
variables and depend on sites.  The site dependence of the conditions means
the unitarity of the time evolution operator \( U(x) \)  cannot be satisfied
by simple algebraic relations between \( A \) and \( B \) in general.
\newline

\noindent
{ \bf 1+1 dimensional case }
\newline

In 1+1 dimensions the relations reduce to the 
previous ones owing to the uni-dimensionality of space.
The unitarity of the time evolution operator is kept
by the same choice of \( A \) and \( B \) as in the free case.
So the path integral formulation is well established keeping the unitarity
at a finite lattice spacing.
\newline

\noindent
{ \bf 1+D dimensional case with $ \bf D \ge 2 $ }
\newline

We cannot determine \( A \) and \( B \) keeping the unitarity
of the time evolution operator in this case.
If we adopt the relations (\ref{u-condition}) for \( A \) and 
\( B \) in the free case,
we obtain from the two relations (3.7) and (3.8)

\begin{eqnarray}
( A_{i}A_{j}^{\dagger} - B_{j}B_{i}^{\dagger} )
    \{ V(x,x+\hat{i})V(x+\hat{i},x+\hat{i}-\hat{j}) 
\makebox[2cm]{} \nonumber \\
- V(x,x-\hat{j})V(x-\hat{j},x+\hat{i}-\hat{j}) \} = 0 , \makebox[7mm]{} ( i \ne j )
\end{eqnarray}

\begin{eqnarray}
( A_{i}B_{j}^{\dagger} - A_{j}B_{i}^{\dagger} )
    \{ V(x,x+\hat{i})V(x+\hat{i},x+\hat{i}+\hat{j})
\makebox[2cm]{} \nonumber \\
- V(x,x+\hat{j})V(x+\hat{j},x+\hat{i}+\hat{j}) \} = 0 .
\end{eqnarray}

We can see the origin of the unitarity violation comes from
the difference between clockwise and anti-clockwise parallel transports
around a plaquette.
This factor is higher order in the lattice spacing \( \sigma \) and expected to be
harmless in the continuum limit.
\newline

\section{Discussion on unitarity}

In Sec.2 we imposed the unitarity condition (\ref{unitary}) and the Klein-Gordon condition
(\ref{Klein}) on the time evolution operator $U$ of the spinor field
\begin{equation}
U=\sum_{i=1}^{D}
    (A_{i} S_{i}
    +B_{i} S_{i}^{\dagger})
    +C.
\label{TEM}
\end{equation}
We could not find the spinor field on discrete space-time with the
same dimension as one on continuous space-time. 

In this section we try to find the spinor field with lower dimension than
the one in Sec.2 without imposing the unitarity condition, since the Klein-Gordon
condition ensures that the spinor field has no doubling.

As in Sec.2 we assume the Klein-Gordon equation:
\begin{equation}
U^{-1}-2+U=r\;\sum^{D}_{i=1}
          (S_{i}-2+S_{i}^{\dagger})
          -\tau^2M^2,
\label{kg}
\end{equation}
where we supposed the existence of inverse matrix for the time evolution
operator. 
From the above equation we can see that the inverse matrix $U^{-1}$ is of 
the same form as  $U$ :
\begin{equation}
U^{-1}=\sum^{D}_{i=1}
     (\bar{A}_{i} S_{i}
     +\bar{B}_{i} S_{i}^{\dagger})
+\bar{C}.
\label{DinvU}
\end{equation}
Generally the term proportional to $(S_{i}^{(\dagger)})^n\;(n > 1)$  may be
included in $U^{-1}$,
but these terms are not allowed, because in Eq. (\ref{kg}) there is 
no term which cancels out those  terms.

From Eq. (\ref{kg}) we obtain the following conditions:
\begin{eqnarray}
A_{i}+\bar{A}_{i} &=&r ,               \nonumber\\
B_{i}+\bar{B}_{i} &=&r ,               \label{kg2:2}\\
C    +\bar{C}     &=&K ,               \nonumber
\end{eqnarray}
where $K$ is 
\begin{equation}
K=2(1-rD)-\tau^2M^2,
\label{constK}
\end{equation}
and from $UU^{-1}=1$ we obtain
\begin{equation}
\begin{array}{ll}
A_i \bar{A}_j+A_j\bar{A}_i=0, &  \mbox{(for all $i$, $j$)}  \\
B_i \bar{B}_j+B_j\bar{B}_i=0, &  \mbox{(for all $i$, $j$)}  \\
A_i \bar{C}     +C     \bar{A}_i=0, &  \mbox{(for all $i$)} \\
B_i \bar{C}     +C     \bar{B}_i=0, &  \mbox{(for all $i$)} \\
A_i \bar{B}_j+B_j\bar{A}_i=0, &  \mbox{$(i \neq j)$}      \\
\sum^{D}_{i=1}(A_i\bar{B}_i+B_i\bar{A}_i)
              +C     \bar{C}=1.         &                \\
\end{array}
\label{inv:2}
\end{equation}

We find that the matrix $(C-K/2)^2$ commutes 
with $A_i$, $B_i$ and $C$, and we can regard this matrix as the unit matrix up 
to a constant factor $DQ$:
\begin{equation}
(C-\frac{K}{2})^2=DQ \openone.
\label{CC}
\end{equation}
Using Eqs. (\ref{kg2:2}) $\sim$(\ref{CC}) and assuming the isotropy of
spatial lattice directions we can get the anticommutation 
relations between
$A_i$, $B_i$ and $C$ 
\begin{equation}
\{M_a,M_b\}=F_a \delta_{a,b}\openone,
\label{antic}
\end{equation}
where
\begin{equation}
F_a=\left\{\begin{array}{lr}
-2\left(\frac{K^2}{4D}-\frac{1}{D}-Q\right), &(  1 \leq a \leq    D) \vspace{10pt} \\
 2\left(\frac{K^2}{4D}-\frac{1}{D}-Q\right), &(D+1 \leq a \leq 2D-1) \vspace{10pt} \\
\lambda_{+}                                , &(a=2D)                 \vspace{10pt} \\
\lambda_{-}                                , &(a=2D+1)              \vspace{10pt} \\
           \end{array} 
    \right.
\label{Fa}
\end{equation}
and $\lambda_{\pm}$ are eigenvalues of the 
$2\times 2$ matrix ${\cal M}$,
which is defined by anticommutation relation between $C-K/2$ and
$\sum_{i=1}^D(A_i+B_i)/\sqrt{D}$
\begin{eqnarray}
{\cal M}&=&\left(\begin{array}{cc}
  \{\sum_{i=1}^D\frac{(A_i+B_i)}{\sqrt{D}}
   ,\sum_{i=1}^D\frac{(A_i+B_i)}{\sqrt{D}}\}  &
  \{\sum_{i=1}^D\frac{(A_i+B_i)}{\sqrt{D}},C-\frac{K}{2}\}  \vspace{10pt} \\
  \{C-\frac{K}{2},\sum_{i=1}^D\frac{(A_i+B_i)}{\sqrt{D}}\}                   &
  \{C-\frac{K}{2},C-\frac{K}{2}\}                           \vspace{10pt}  \\
               \end{array} \right) \vspace{10pt} \nonumber\\
&=&\left(\begin{array}{cc}
            2(Dr^2-Q+\frac{K^2-4}{4D})  &\sqrt{D}Kr \vspace{10pt} \\
            \sqrt{D}Kr                  & 2DQ       \\
         \end{array}
   \right). 
\end{eqnarray}
Clearly $\lambda_{\pm}$ are real since the matrix ${\cal M}$ is 
hermitian .
The matrices $M_a$'s are defined by
\begin{equation}
M_a=\left\{\begin{array}{lr}
A_{a}-B_{a},                &(1 \leq a \leq D)     \vspace{10pt} \\
    \sum_{i=1}^D\xi^{i}_{a-D}(A _{i}+B_{i}), 
                           &(D+1 \leq a \leq 2D-1)\vspace{10pt} \\
    \eta^{(1)}_{+}\sum_{i=1}^D\xi^{i}_{0}(A _{i}+B _{i})
   -\sqrt{D}r+\eta^{(2)}_{+}(C-\frac{K}{2}),
                           &(a=2D)                \vspace{10pt} \\
    \eta^{(1)}_{-}\sum_{i=1}^D\xi^{i}_{0}(A _{i}+B _{i})
   -\sqrt{D}r+\eta^{(2)}_{-}(C-\frac{K}{2}),
                           &(a=2D+1)                            \\
            \end{array}
   \right.
\end{equation}
where $\eta^{(\rho)}_{\pm}$ are the eigenvectors of ${\cal M}$ with 
eigenvalues $\lambda_{\pm}$:
\begin{equation}
{\cal M}\left(\begin{array} {c}\eta^{(1)}_{\pm} \\
                               \eta^{(2)}_{\pm} \\
             \end{array}
       \right)=\lambda_{\pm}
       \left(\begin{array} {c}\eta^{(1)}_{\pm} \\
                              \eta^{(2)}_{\pm} \\
             \end{array}
       \right),
\end{equation}
and the ${\xi^{i}_{\alpha}}$'s, $(\alpha=0,1,...,D-1)$ are 
orthogonal vectors:
\begin{equation}
\sum_{i=1}^{D}\xi^{i}_{\alpha}\xi^{i}_{\beta}=\delta_{\alpha,\beta}.
\end{equation}
We do not give the definite form of these vectors except for 
\begin{equation}
\xi^{i}_{0}=\frac{1}{\sqrt{D}}
            \left(\begin{array}{c}  
                    1 \\
                    1 \\
                \cdot \\
                \cdot \\
                \cdot \\
                    1 \\
                   \end{array}
             \right),
\end{equation}
since we do not need those forms in the later discussion.

From these equations we obtain
\begin{equation}
\{M^{\dagger}_a,M^{\dagger}_b\}=F_a \delta_{ab}{\openone}.
\label{antcdag}
\end{equation}
There is no condition which prescribes the anticommutation relations 
between $M_a$ and $M^{\dagger}_a$ . 
For simplicity we assume the following anticommutation 
relation:
\begin{equation}
\{M       _a,M^{\dagger}_b\}=D_{ab}{\openone}.
\label{antcttl}
\end{equation}

Noticing that $M_a$ tends to a linear combination of $\gamma$
 matrices in continuous space-time limit, we can easily see that the case with 
 $Q=(K^2/4-1)/D$ is 
undesired.  In this case all matrices $M_a$  vanish in the continuous 
space-time limit except for the two matrices $M_{2D}$ and $M_{2D+1}$, although we need 
$D+1$ matrices in order that the equation of spinor field goes to Dirac 
equation in the limit $\sigma, \tau \;\;\rightarrow\;\; 0$.

We consider the case with $F_a \neq 0\;\;(a=1,...,2D+1)$.
It is convenient to use 
\begin{eqnarray}
\tilde{M}_a&=&\frac{M_a}{\sqrt{F_a}}\nonumber\\
           &=&\tilde{X}_a+ i\tilde{Y}_a. 
\end{eqnarray}
The anticommutation relations between these matrices become
\begin{eqnarray} 
\left\{\;\; 
     \left(
            \begin{array}{c}
              \tilde{X}_{a} \\ 
              \tilde{Y}_{a} 
            \end{array}
     \right)
,
     \left(
            {\tilde{X}_{b},\tilde{Y}_{b}} 
     \right)
\right\}=
\left(
\begin{array}{cc}
      \delta_{ab}+{\cal{G}}_{ab}  & {\cal{F}}_{ab} \\
        -{\cal{F}}_{ab}           & {\cal{G}}_{ab}
\end{array}
\right),
\label{AccNorXY}
\end{eqnarray}
where
\begin{eqnarray}
{\cal G}_{ab}
&=&-\frac{1}{2}(\delta_{ab}-\tilde{D}_{ab}-\tilde{D}^{\ast}_{ba}){\openone},
              \nonumber\\
{\cal F}_{ab}
&=&\frac{1}{2 i}(\tilde{D}_{ab}-\tilde{D}^{\ast}_{ba}){\openone},
\label{antcxy}
\end{eqnarray}
and $\tilde{D}_{ab}$ is 
\begin{equation}
\tilde{D}_a=\frac{D_{ab}}{\sqrt{F_aF_b}}.
\nonumber
\end{equation}
As the eigenvalues of matrix ${\cal G}_{ab}$ are non-negative, the matrix
$\delta_{ab}+{\cal G}_{ab}$ is invertible. Using the inverse matrix of 
$\delta_{ab}+{\cal G}_{ab}$, we define 
\begin{equation}
\tilde{Y}'_{a}=\sum_{b=1}({\cal F}(\openone+{\cal G})^{-1})_{ab}X_b+Y_a,
\end{equation}
and then we have
\begin{eqnarray} 
\left\{\;\; 
     \left(
            \begin{array}{c}
              \tilde{X}_{a} \\ 
              \tilde{Y}'_{a} 
            \end{array}
     \right)
,
     \left(
            {\tilde{X}_{b},\tilde{Y}'_{b}} 
     \right)
\right\}=
\left(
\begin{array}{cc}
      \delta_{ab}+{\cal G}_{ab} & 0 \\
        0           & {\cal D}_{ab}+
                     ({\cal F}(\openone+{\cal G })^{-1}{\cal{F}})_{ab}
\end{array}
\right).
\end{eqnarray}
As the upper-left submatrix of the right hand side of the above equation is 
invertible, we cannot make the number of independent matrices $X_a$ and
$Y_a$ smaller than 2D+1. Thus the dimension of these matrices is larger than 
$2^D$( for example, the lower limit of the dimension is 8 for $D=3$). 

If $\lambda_{+}$ and $\lambda_{-}$ are zero, the dimension of matrices
is equal to or larger than $2^{D-1}$ . If either $\lambda_{+}$ or $\lambda_{-}$ 
is zero, the dimension of matrices is equal to or larger than $2^D$. For
example, the limit of the dimension is 4 or 8 for $D=3$. 
The massless spinor field in Sec.2 corresponds to the former case.

In conclusion  we found under the assumption (\ref{antcttl}) that the dimension of spinor field without the 
unitarity condition is  not smaller than the dimension with the condition.


\section{Summary}

We have formulated a free fermion without doubling on 1+D dimensional Minkowski 
lattice space-time.  We required there the unitarity of time evolution operator 
and Klein-Gordon equation on lattice space-time.  This means the norm is 
conserved and the fermion has no doubler.
We showed that the minimal number of components of massless field $\Psi$ is
\( 2^{\em D-1} \).  In 1+1 dimensional case the equation is the same as that of
the cellular automaton by 't Hooft et al.  In 1+3 dimensional massless case
the time evolution operator was expressed in an explicit form using usual 
$\gamma$ matrices, which tends to 
Dirac operator in the continuous space-time limit.  

The action is not hermitian.  We proved the equivalence of canonical
quantization and that through the path integral.
We have given the explicit form of the fermion propagator,
which has no extra poles of doublers.  

In the case where the fermion interacts with gauge fields the action was also
written in a gauge invariant form. 
The vacuum expectation value of an arbitrary
operator is calculated by the path integral formulation in principle.
The time evolution operator is not unitary in this case,
and the reason was considered briefly.  

We have tried to find the spinor field with lower dimension than $2^{\em D-1}$
without imposing the unitarity condition.
However, we found the spinor field with the smallest dimension 
is $2^{\em D-1}$ under a certain condition.


\end{document}